\documentclass[12pt,preprint]{aastex}

\newcommand{\kms}{\hbox{km s$^{-1}$}}

\shorttitle{[Ne II] Emission from TW Hya}
\shortauthors{Herczeg et al.}

\begin{document}

\title{High-resolution Spectroscopy of [Ne II] Emission from TW Hya}

\author{Gregory J. Herczeg\altaffilmark{1}, Joan
  Najita\altaffilmark{2}, Lynne A. Hillenbrand\altaffilmark{1}, Ilaria Pascucci\altaffilmark{2}}

\altaffiltext{1}{Caltech, MC105-24, 1200 E. California Blvd., Pasadena, CA 91125}
\altaffiltext{2}{National Optical Astronomy Observatory, Tucson, AZ 85719}
\altaffiltext{3}{Steward Observatory, The University of Arizona, Tucson, AZ 85721}

\begin{abstract}
We present high-resolution echelle spectra of [\ion{Ne}{2}]
$12.81\mu$m emission from the classical T Tauri star (CTTS) TW Hya
obtained with MICHELLE on Gemini North.  The line is centered at the
stellar radial velocity and  
has an intrinsic FWHM of $21\pm4$ \kms.
The line width is broader than other narrow emission lines typically
associated with the disk around TW Hya.  If formed in a disk, the line broadening could result from
turbulence in a warm disk atmosphere, Keplerian rotation at an average
distance of 0.1 AU
from the star, or a photoevaporative flow from the optically-thin region
of the disk.
We place upper limits on the [\ion{Ne}{2}] emission flux from the CTTSs DP Tau and BP Tau.
\end{abstract}
 
\keywords{
  stars: pre-main sequence --- stars: planetary systems:
  protoplanetary disks matter --- infrared: stars: TW Hya, DP Tau, BP Tau}

%
%


\section{INTRODUCTION}
X-ray and UV emission from classical T Tauri stars (CTTSs) can affect the chemical structure of their protoplanetary disks by ionizing the disk surface, dissociating molecules, and heating the gas and grains \citep[see reviews by][]{Naj07,Dul07,Ber07}.   
Direct evidence of irradiated disks is seen in far-UV H$_2$ emission, which requires excitation by \ion{H}{1} Ly$\alpha$ irradiation and can be attributed to emission from the inner disk of some CTTSs \citep{Her06}.  Other observational evidence for FUV- and X-ray-irradiated disks around CTTSs has been more circumstantial.  
The underabundance of HCN relative to CN in disks around several CTTSs
may be attributed to strong Ly$\alpha$ emission \citep{Ber03}.
\citet{Bar03} suggest that X-ray irradiation warms the disk surface
enough to produce emission in the H$_2$ 1-0 S(1) line.  The excess FUV
continuum emission may result from the electronic cascade and
dissociation following excitation of H$_2$ by energetic electrons
\citep{Her04,Ber04}.  EUV irradiation of the disk can produce a photoevaporative wind, which could be visible in the low-velocity component of some optical forbidden lines \citep[see review by][]{Dul07}.

Recently, \citet{Pas07} and \citet{Lah07} may have found a new probe of gas at the disk surface by detecting
emission in the [\ion{Ne}{2}] $12.81\mu$m fine-structure line in
low-resolution {\it Spitzer} IRS spectra of CTTSs.  [\ion{Ne}{2}] was
detected in four of six targets in the \citet{Pas07} sample and in 15 of 76 targets in the \citet{Lah07} sample.  Relative to the $\sim 1$ Myr old CTTSs in Taurus, the
 six stars observed by \citet{Pas07} have faint mid-IR continuum
 emission, are X-ray bright, and have mass accretion rates an order of
 magnitude lower than the average CTTS of similar spectral type.  The
 \citet{Lah07} sample and detections are comprised mostly of T Tauri stars with an IR
 excess.  The only class I source in their sample was also detected in
 [\ion{Ne}{2}],
 and several Herbig Ae/Be stars were undetected in [\ion{Ne}{2}].
 The sources both with and without [\ion{Ne}{2}] detections from \citet{Lah07}
 show a range of X-ray luminosities.
  \citet{Lah07} also detected [\ion{Ne}{3}] emission from one source (Sz102),
 and four of the six targets with H$_2$ 0--0 S(2) emission also show [\ion{Ne}{2}] emission.

\citet{Pas07} and \citet{Lah07} both argue that the irradiated disk surface is the most likely
origin for this emission.  Two ionization paths can ionize Ne and thereby
produce the [\ion{Ne}{2}] emission.  \citet{Gla07} propose that the
line is formed because of Ne ionization by K-shell absorption of
stellar X-rays at energies $>0.9$ keV.  The observed [\ion{Ne}{2}]
fluxes and the [\ion{Ne}{3}]/[\ion{Ne}{2}] flux ratio and lower limits
are consistent with predictions from this X-ray ionization model.  
Alternatively, EUV photons shortward of the \ion{Ne}{1} ionization edge
at $575$ \AA\ can produce [\ion{Ne}{2}] emission if the EUV emission
is able to penetrate through neutral gas in the accretion column and
wind.  
If the disk origin is confirmed, then the [\ion{Ne}{2}] 12.81$\mu$m line would be a direct probe of the irradiation of the protoplanetary disk by the central star. 

In this paper, we present high-resolution echelle spectra of the 12.81$\mu$m
region of the CTTSs TW Hya, BP Tau, and DP Tau obtained with MICHELLE
on Gemini North.  TW Hya is a nearby 10 Myr old CTTS with a disk that
is believed to be optically-thin within 4 AU from the star based on
models of its infrared spectral energy distribution \citep{Cal02} and
an inner hole and bright ring imaged at 7mm with the VLA \citep{Wil07}.  BP Tau and DP Tau are 1 Myr old CTTSs with optically-thick inner disks \citep{Ken95,Muz03}.  All three stars are X-ray bright (Table 1).  
We find strong
[\ion{Ne}{2}] emission from TW Hya and place upper limits on the
[\ion{Ne}{2}] flux from DP Tau and BP Tau.  
We discuss the origin of the
emission in the context of profiles of other emission lines from TW
Hya and suggest that the line is formed in a disk, although formation
in a heated photosphere cannot be ruled out.
If the  [\ion{Ne}{2}] emission is formed in a disk, the line width could result from Keplerian
rotation at a distance of 0.1 AU from the star, tran-sonic turbulence,
or detecting the front and back of a photoevaporative flow.

\section{OBSERVATIONS AND DATA REDUCTION}
We used the echelle mode of MICHELLE on Gemini North to observe the
CTTSs TW Hya, DP Tau, and BP Tau on 9 March 2007.  MICHELLE has a
320x240 Si:As array with a plate scale of $0\farcs201$ \citep{Gla97}.  When used
with the echelle
spectrograph MICHELLE has a spectral resolution of $\sim 30,000$.
We centered the targets on the slit using images obtained with the
Si-5 filter, which has an effective wavelength of $11.7$ $\mu$m.  We used a 2-pixel wide slit and the Si-6 filter to obtain spectra centered at 
12.815$\mu$m and covering $\sim0.049$ $\mu$m.  We nodded
$10^{\prime\prime}$ off the slit at 35s intervals.  Our total
on-source integration was 40.4 min for TW Hya, 20 min for BP Tau and
22.5 min for DP Tau.  These three targets were observed at median
airmasses of 1.8, 1.3, and 1.9, respectively.  Observations of
$\alpha$ CMa and the asteroid Iris were used as telluric standards.
The seeing varied from $0\farcs6$ to $0\farcs85$, as measured in the cross-dispersion spatial profile from our unresolved telluric standards.  

In each set of four nod positions, the counts in each column of the off-source images are scaled to and then subtracted from the counts in the same column of the on-source sky background.  This scaling is necessary because the second derivative of the sky brightness versus time was non-zero. 
We correct for fringing in the cross-dispersion direction by fitting a high-order polynomial to the residual background in each column.  
We extract the spectrum from a 7-pixel window centered on the source,
and then subtract background sky spectra obtained from 20-pixel
windows on both sides of the source.  Fringing in the spectral
direction is not detected or corrected for because of the low S/N in
the continuum (Table 1).  

We estimate the telluric correction for each observation by
interpolating the standard spectra of Iris and $\alpha$ CMa at
intermediate airmasses.  The atmospheric correction is applied to each
nod prior to combining the images.  Several O$_3$, CO$_2$, and H$_2$O
lines are present in the wavelength range of our spectra.
We calibrate wavelengths to an accuracy of $\sim 2$ \kms\ per pixel
using a 2nd-order fit to sky emission lines identified with ATRAN
\citep{Lor92}.  Each pixel covers $\sim 3.6$ \kms.  We estimate a
resolution of $3.2\pm0.5$ pixels ($11.5\pm1.8$ \kms) from the FWHM of
four weak, narrow sky emission lines.

 TW Hya was also observed on 12 April 2006 during re-commissioning of the East Arm Echelle
Spectrograph \citep{Lib95} on the 5m Hale Telescope at Palomar.  The
instrument is fed by a bundle of fibers that together subtend
$1\farcs5$ and illuminate a $2048\times2048$ pixel CCD.  The spectrum
covers most of the 4000-9000 \AA\ range at a resolution of
15,000-45,000, depending on location on the chip.  The relative wavelength solution in each order was obtained using a ThAr lamp.  The absolute
wavelength solution is determined separately for each echelle order by matching photospheric lines to the radial velocity of TW Hya and is accurate to $\sim 3$ \kms.

No sufficient flat field was obtained during the commissioning.
However, observations of the white dwarf Feige 34 and the A0 star
HD74198 obtained in the same night yield spectra with a count rate
that peaks near the center of each order and falls off smoothly to the
edge of the detector.  The absence of flat-fielding has  a negligible effect on the
line profile shapes and equivalent width measurements because they are
extracted from narrow spectral regions.  
We use observations of several other stars to identify the approximate flux in telluric emission lines, which are located at -20 \kms\ relative to the photospheric velocity of TW Hya.  
The telluric and stellar components of the [\ion{O}{1}] lines at
5577, 6300, and 6363 \AA\ can therefore be separated.  The \ion{Na}{1} D emission lines are undetected, with a sensitivity
limited by strong telluric emission.  
We calculate line fluxes from equivalent widths by estimating the
continuum flux level from non-simultaneous archival {\it HST}/STIS
optical spectra \citep{Her04,Rob05}.  

The heliocentric radial velocity of TW Hya is $12.2\pm0.5$ \kms\
(Weintraub et al. 2000; see also Kastner et al. 1997, de la Reza et
al. 1989) and of BP Tau is 15.8 \kms\ \citep{Her95}.  The heliocentric
radial velocity of DP Tau has not been measured.  We estimate a radial
velocity of $17\pm5$ \kms\ by averaging the radial velocities of the T Tauri stars GO Tau and V955 Tau \citep{Her95}, which are both about 10$^\prime$ from DP Tau.

\section{CHARACTERIZING THE [\ion{Ne}{2}] EMISSION}
Figure 1 shows our MICHELLE spectra of TW Hya, BP Tau, and DP Tau.  The
continuum is detected from each target.  We measure equivalent widths
for the [\ion{Ne}{2}] fine-structure line ($^2P_{3/2}$ -- $^2P_{1/2}$)
at $12.81355\pm0.00002$ $\mu$m \citep{Yam85} of $62\pm11$ \AA\ for TW Hya and $2\sigma$ upper limits of $<10$ and $<11$ \AA\ for non-detections from BP Tau and DP Tau, respectively.  The upper limits are calculated by assuming the same line profile as that of TW Hya, and would be larger if the hypothetical lines from these objects were broader.
We convert equivalent widths to $(5.9\pm1.1)\times10^{-14}$, $<0.7\times10^{-14}$, and $<1.1\times10^{-14}$ erg cm$^{-2}$ s$^{-1}$ using $12.8\mu$m continuum flux measurements  from {\it Spitzer} IRS spectra \citep{Uch04,Fur06}.  These
fluxes assume that the mid-IR continuum and the [\ion{Ne}{2}] emission are not
spatially extended beyond the seeing. 

The [\ion{Ne}{2}] line profile from TW Hya is consistent with a single Gaussian profile
centered at $-2\pm3$ \kms\ in the stellocentric frame with a FWHM of $6.7\pm0.8$ pixels ($24\pm3$ \kms).
Deconvolving the line width from the instrumental resolution yields an
intrinsic line FWHM of $21\pm4$ \kms.  Two components (a broad
component at the stellocentric velocity
with a flux enhancement on the blue side of the line)
are visually apparent in the line profile.  However, based on the S/N
detected in the continuum (Table 1) the profile shape is not a significant deviation from a single Gaussian profile.  
The telluric CO$_2$ line did not divide out sufficiently for some of
the TW Hya observations and results in increased uncertainty on the blue side of the line profile.  The 
[\ion{Ne}{2}] emission is more compact than the $0\farcs75$ angular
resolution (40 AU at the distance of TW Hya), which was estimated from both the mid-IR continuum of TW Hya and the telluric
standard $\alpha$ CMa, which was observed shortly before TW Hya.

Table 2 and Figure 2 compare the [\ion{Ne}{2}] emission to a sample of other emission lines from TW
Hya, excluding those significantly affected by wind absorption (e.g.,
\ion{H}{1} $\lambda6563$, \ion{O}{1}
$\lambda 1302$, \ion{C}{2} $\lambda1335$, \ion{He}{1}
$\lambda10830$, see Herczeg et al. 2002, Edwards et al. 2006).  
These lines show a range of emission profile shapes, and were observed at resolutions similar to that of our MICHELLE spectra.
The velocity and FWHM listed in Table 2 are measured from Gaussian fits to the emission lines.
The energy of the upper level ($E^\prime$) for these transitions
(0.60--23.1 eV) are all substantially larger than that of the
fine-structure [\ion{Ne}{2}] line (0.10 eV).  
Most lines, including the \ion{Ca}{2} IR triplet and the 
 [\ion{O}{1}] $\lambda6300,6363$, show narrow emission centered at
the stellocentric velocity.  No high-velocity ($>100$ \kms) or
low-velocity ($<20$ \kms) component of other optical forbidden lines
are detected \citep{Ham94,Har95}.  Several lines show broad
redshifted emission profiles, presumably produced by the accretion
flow.  Many of the near-UV \ion{Fe}{2} lines have both narrow and broad
components centered at the stellocentric velocity.

\section{DISCUSSION}

\subsection{Comparison To Other Emission Lines From TW Hya}
Emission lines from CTTSs may be formed at or near the disk surface,
in the accretion funnel flow and shock, in outflows, in the stellar
chromosphere, or in the accretion-heated part of the stellar photosphere.  
We first rule out formation of the [\ion{Ne}{2}] emission in the accretion
flow and shock.   Lines commonly attributed to accretion, including \ion{He}{1} $\lambda5876$ and the
\ion{C}{4} $\lambda1549$, \ion{N}{5} $\lambda1240$, and \ion{O}{6}
$\lambda1035$ doublets \citep{Ale02,Her02,Lam04,Joh07,Yan07}, have much
broader line widths than the [\ion{Ne}{2}] emission from TW Hya and 
are redshifted from the stellocentric velocity.  Second,
we rule out fast outflows because the [\ion{Ne}{2}] line is centered
at or near the stellocentric velocity and is limited to low velocities.  Formation in a very slow
outflow may be possible if both the front and back sides of the flow
are visible.  Third, we rule out a
chromospheric origin.  Some lines from Table 2 are attributed to
chromospheric emission and seen from main-sequence dwarfs and
weak-line T Tauri stars \citep{Ayr05}, but these lines are
narrower than the [\ion{Ne}{2}] line. 
 Several other emission lines, including the \ion{C}{1} $\lambda1656$
 and \ion{C}{2}]
 $\lambda2326$ multiplets, \ion{Cl}{1} $\lambda1351.7$, and some NUV
 \ion{Fe}{2} lines, have slightly larger 
 widths than the [\ion{Ne}{2}] emission and may form in a
 heated stellar photosphere.  However, these lines have larger energy upper
 levels than the fine-structure
 [\ion{Ne}{2}] line and are unlikely to trace the same gas.
More likely is the possibility that the [\ion{Ne}{2}] line forms 
either in the inner disk or in a photoevaporative flow from the 
optically thin region of the disk, as suggested by 
the excitation energy (0.10 eV), width (21 \kms), and 
velocity centroid ($-2\pm3$ \kms) of the line. 


Collisional ionization equilibrium would require 
temperatures of $>30,000$ K to produce [\ion{Ne}{2}] emission.  
All identified lines from TW Hya that probe $>20,000$ K have a larger width
($>40$ \kms) than the [\ion{Ne}{2}] line.  We therefore infer that the
[\ion{Ne}{2}] emission is not likely collisionally ionized.
Instead, the emission is likely ionized
by X-ray or EUV-irradiation of gas at or near the disk surface.   
Indeed, Glassgold et al.~(2007) speculated that X-ray irradiated 
disks would produce strong [\ion{Ne}{2}] emission.  The observed 
width of the [\ion{Ne}{2}] line is larger than that shown by 
Glassgold et al.~(in their Fig.~5).  However, Glassgold et al.~(2007) 
considered only the emission that would arise from a non-turbulent 
X-ray irradiated disk that extends beyond 1\,AU.

Irradiated disks are believed to produce the molecular line emission 
detected from TW Hya in the FUV H$_2$, near-IR H$_2$, 
and fundamental CO transitions 
\citep{Wei00,Her02,Ret04,Sal07,Naj07}.
The emission arises from warm ($500-3000$ K) gas, and the
FUV H$_2$ lines in particular require irradiation by strong FUV
emission from the central star.  The line widths are $>16$, $8$, and $14$ \kms\ narrower for H$_2$ 1-0
S(1), FUV H$_2$, and fundamental CO lines, respectively, than the
[\ion{Ne}{2}] line.  The narrow width of the H$_2$ 1-0 S(1) suggests
formation at large radii.  

Interestingly, the widths of the 
fundamental CO, FUV H$_2$, and [\ion{Ne}{2}] lines 
increase with their approximate formation temperatures of 
$\sim 500-1500$\,K (Salyk et al. 2007; Najita et al.\ 2003), 
$\sim 2500$\,K (as observed for TW Hya; Herczeg et al.\ 2004), 
and 4000--10,000\,K (based on theoretical modeling; 
Glassgold et al.\ 2007; Hollenbach et al.\ 2000), 
respectively. 
This trend might indicate a common origin for the lines in a disk 
with an outwardly decreasing temperature gradient (e.g., Carr et al.\ 2004). 
If the [\ion{Ne}{2}] line width results from
Keplerian disk rotation, then the emission arises at an average
distance of $\sim 0.1$
AU of the star, assuming a $7\pm1^\circ$ inclination of the disk
around TW Hya \citep{Qi04}.  
The trend might also indicate a common origin for the lines in a turbulent 
disk atmosphere with a vertically increasing temperature gradient,- i.e.,  
increasing (transonic) turbulent broadening higher in the disk 
atmosphere.  For example, 
the line width of the [\ion{Ne}{2}] emission could be explained 
entirely by turbulent broadening in 10,000 K gas.
With a combination of Keplerian rotation and turbulence, the 
line could form at disk radii larger than $\sim 0.1$ AU and in gas
cooler than $10,000$ K.
Alternatively, 
the width of the [\ion{Ne}{2}] line could also be explained 
if the emission arises from a photoevaporative flow from the 
optically thin region of the disk (at $<4$ AU).
We explore these possibilities using simple scaling arguments to 
point out the challenges associated with these interpretations.

\subsection{X-ray Excitation of the Inner Disk}

The [\ion{Ne}{2}] flux is converted to a mass from the Einstein
A-value of $8.59\times10^{-3}$ s$^{-1}$ \citep{Kra06}, a solar Ne/H abundance
ratio of $8.9\times10^{-5}$ \citep{And89}, a Ne ionization
fraction of $\zeta$, and an average atomic mass per atom of 1.4.  In LTE the fraction of \ion{Ne}{2} in the
$^2P_{3/2}$ level is 0.14 and 0.28 at 1000 and 5000 K if the electron
density is larger than the critical density, $10^5$ cm$^{-3}$, as
assumed here.  If the line is optically-thin and not attenuated by
dust within the disk, then the gas mass is
$0.5-1.5\times10^{-9}{\zeta^{-1}}$ $M_\odot$, where this range
includes uncertainties in the line flux and the population of the upper level.  If we assume that the
[\ion{Ne}{2}] emission arises in the annulus of an empty cylinder with
a height $H$, 
radius $R$, and width $dR<R$, then the total hydrogen column density of the layer
measured parallel to the disk is $N($H$)\sim 1.5\times
10^{22}\frac{0.2 AU}{R}\frac{0.2 AU}{H}  \zeta^{-1}$ cm$^{-2}$. 

Either the disk atmosphere or the photoevaporative flow scenario requires that the ionizing
photons penetrate through the emission region, which 
constrains the neutral hydrogen column density, $N$(\ion{H}{1}), as
measured radially from the central star.
Based on the {\it Chandra} X-ray spectrum of TW Hya \citep{Kas02},
23\% of photons with energies $>0.9$ keV penetrate to depths of
$N$(\ion{H}{1})=$10^{22}$ cm$^{-2}$, but only the hardest 3\% of those
photons can penetrate to $N$(\ion{H}{1})=$10^{23}$ cm$^{-2}$.
However, almost all EUV photons longward of 100 and 500 \AA\ are attenuated
for $N$(\ion{H}{1})$>10^{20}$ and $10^{18}$ cm$^{-2}$, respectively.

If we set $R\lesssim0.2$ AU, then $N$(H)$\sim 10^{22} \zeta^{-1}$
cm$^{-2}$.  Even if the ionization fraction is high, $N$(\ion{H}{1})
will be large enough to require the ionization of
Ne by X-rays and a disk height comparable to the disk radius.
Such a  
geometrically thick disk may also be required to explain the strength
of the H$_2$ fluorescence in FUV spectra of TW Hya \citep{Her04}.
The scale height of gas in hydrostatic equilibrium at $R=0.2$ AU from
TW Hya can be estimated as $0.04 \left(\frac{T}{10^4 K}\right)^{-0.5}$
AU.  \citet{Eis06} calculated that the dust temperature is $\sim1100$
K at the inner dust truncation radius ($\sim 0.06$ AU).  However,
since only a small amount of micron-sized dust is present within the 4
AU dust clearing radius \citep{Cal02}, the gas temperature at
these radii may be significantly hotter than the dust temperature. 
Models of optically-thick disks with viscosity parameter $\alpha=0.01$ suggest a gas
surface density of 25--90 g cm$^{-2}$ at 0.1 AU from TW Hya
\citep{Dal98}. The mass of the [\ion{Ne}{2}] emitting region would be
about $\frac{0.1}{\zeta}$\% of the total disk mass within 0.2 AU of
the star.  The [\ion{Ne}{2}] emission in this scenario traces a thin
surface layer of the disk higher than one scale height above the disk
midplane.  This interpretation of [\ion{Ne}{2}] emission is
challenging if $\zeta<0.1$ because 
the larger total mass traced by the
[\ion{Ne}{2}] emission would require an even larger geometrical
disk height.

This situation can be ameliorated if turbulent broadening contributes
to or dominates the [\ion{Ne}{2}] line width.  If the disk atmosphere
is turbulent and the turbulent velocity is close to the sound speed,
then the FWHM of a line would be $\sim 21$ \kms\ at $10,000$ K,
consistent with the observed line width.  At this temperature the
[\ion{Ne}{2}] emission could arise from a wide range of radii, perhaps
extending out to 10-20 AU.  \citet{Gla07}
suggest that at 1--10 AU, the gas within $N(H)=10^{20}$ cm$^{-2}$ of the disk
surface will be heated to 4000 K.  At this temperature the average
radii for formation of [\ion{Ne}{2}] emission would be $\sim 0.5$ AU
from the star.

Approximately transonic turbulence in disk atmospheres is inferred
from the intrinsic line broadening of CO overtone emission from
accretion disks (Carr et al. 2004; Najita et al. 1996).  The
dissipation of turbulence in disk atmospheres could also produce the
extra heating that is needed to explain the strength of CO fundamental
emission from T Tauri disks \citep{Gla04}.  While these
observations refer to emission from cool, molecular disk gas that is
located deeper in the disk atmosphere, similar transonic
turbulence may characterize warmer, atomic gas higher up in the disk
atmosphere where the [\ion{Ne}{2}] emission may arise.   Indeed,
simulations of weakly magnetized disks find that MHD turbulence
generates a warm ($\sim 10,000$ K) disk corona with a transonic velocity
dispersion above several disk scale heights \citep{Mil00}.

\subsection{Photoevaporation of an EUV-irradiated Disk Surface}

On the other hand, in the disk dissipation models of Alexander et al. (2006ab)
EUV irradation of the disk produces a $10,000$ K flow with a velocity of
$\sim 10$ \kms\ from the disk surface, near the gravitational radius
$r_g$ (see also Hollenbach \& Gorti 2007).  
Analytical formulations
for $r_g$ assuming an abiabatic gas \citep{Lif03} suggest that
$r_g\approx 1.4\frac{M_*}{M_\odot}\frac{10^4 K}{T}$AU$\approx 1$ AU
for a CTTS with the mass of TW Hya.  In the photoevaporation simulations by \citet{Fon04},
most of the mass loss occurs inside $\sim 6$ AU for a source like TW
Hya.  While these expectations are for a radially continuous gas disk,
\citet{Cal02}
find that at 10 $\mu$m the disk is optically thin within 4 AU of TW
Hya.
Any additional frontal illumination of the inner edge of the optically-thick
disk at 4 AU by FUV and X-ray photons may provide an additional heating
source that could help puff up the disk or help launch the
photoevaporative flow. 
If the [\ion{Ne}{2}] emission is formed in a photoevaporative flow
within 4 AU of the star, the apparent line broadening could result
from detecting  [\ion{Ne}{2}] emission from both the front and back
sides of the disk. 
The profile
could be slightly blueshifted if some [\ion{Ne}{2}] emission occurs in a photoevaporative
flow beyond the dust clearing radius, where emission on the back
side of the disk would not be detectable.  
\citet{Ale06a} predict
that the photoevaporative flow will have velocities $>10$ \kms\ at
distances greater than 4 AU
from the star, implying the possibility of a stronger blueshifted
asymmetry than is observed.


If we assume the photoevaporative flow is produced at $R\sim4$ AU,
then the height of the flow (including both front and back sides of
the disk) is
$H=\frac{1.5\times10^{20} {\rm cm}^{-2}}{\zeta N({\rm H})}$
AU.  If $\zeta=1$ and $N$(H)$\sim10^{19}$ cm$^{-2}$, then $H=15$ AU.  
 The velocity of a photoevaporative flow is expected to be $v\sim
 10$ \kms\ at 4 AU above the disk \citep{Ale06a}.  The crossing time for the evaporating gas across the
 emission region on either side of the disk ($H/2$) is then 3.5 yr.
 The estimated mass loss rate is
$\frac{M_{{\rm Ne II}}}{3.5 {\rm yr}}\sim 3\times 10^{-10} f$ $M_\odot$ yr$^{-1}$, where $f$ is the fraction
of [\ion{Ne}{2}] emission that arises in the photoevaporative flow.

Such a low-velocity ($\sim 10$ \kms) photoevaporative flow would be distinct from the observed
high-velocity ($-50$ to $-200$ \kms) wind
that dominates blueshifted absorption profiles of 
atomic and singly-ionized lines from
TW Hya (Edwards et al. 2006;
Johns-Krull et al. 2007).
The estimated photoevaporation mass loss rate
of about one-fifth the stellar mass accretion rate is similar to the $\sim
3\times10^{-10}$ $M_\odot$ yr$^{-1}$ estimate for a
radially-continuous disk that is irradiated with an 
 EUV photon flux of $\sim10^{41}$ phot s$^{-1}$ \citep{Hol94}.  Several
parameters in this simple scaling argument, including $f$, are uncertain and together could
suppress our estimate for the mass loss rate.  This scenario
constrains the $N$(H) in the photoevaporative flow to a narrow range.
A smaller $N$(H) would increase
the volume and decrease the $n_e$ below the critical density by
increasing $H$, where any flow is likely less collimated.
A larger $N$(H) decreases the height and crossing time, thereby
increasing the estimate of the mass loss rate. 

The photoevaporation scenario is contingent upon some EUV emission
reaching the disk surface.  The photon flux at $<912$ \AA\ from TW Hya
of $\sim 10^{41}-10^{42}$ phot s$^{-1}$ is dominated by line emission
\citep{Ale05}.  Since both the $\sim 10^5$ K FUV lines and $10^6$ K X-ray
lines are attributed to the accretion flow \citep{Joh07,Kas02}, this
estimated EUV emission likely applies to emission produced near the
accretion shock.  Whether this EUV emission reaches the disk depends
on a small neutral hydrogen column density.  A smaller amount of EUV emission is
expected to be produced by the transition region in magnetic
structures and may reach the disk, depending on the geometry of the
accretion flow.
Based on an analysis of FUV H$_2$ emission, \citet{Her04} suggested
that an absorbing column of $N$(\ion{H}{1})$\sim 5\times10^{18}$ cm$^{-2}$
\ion{H}{1}, perhaps in a stellar outflow, is present between the \ion{H}{1} Ly$\alpha$
emission from TW Hya and the warm H$_2$ in the disk.  If $N$(\ion{H}{1})$\lesssim2\times10^{18}$
between the EUV and [\ion{Ne}{2}] emission resions, then
5--15\% of the EUV photons could reach the disk and lead to
some photoevaporation.  A larger intervening $N$(\ion{H}{1}) would
prevent EUV photoevaporation.


\subsection{Conclusions}

Our lower luminosity limits for
[\ion{Ne}{2}] emission from BP Tau and DP Tau are comparable to the
line luminosity from TW Hya, despite having mass accretion rates an
order of magnitude larger than that onto TW Hya.  [\ion{Ne}{2}]
is detected from several CTTSs with low mass accretion rates both here and
in \citet{Pas07}.
Of the few stars in the \citet{Lah07} sample with measured mass
accretion rates, [\ion{Ne}{2}] emission is detected from several weak
accretors but not from several strong accretors.  If weak accretors do
show stronger
[\ion{Ne}{2}] emission than strong accretors, this effect
may be explained because the intervening 
$N$(\ion{H}{1}) in the accretion flow and wind should scale with the mass accretion and mass loss
rates.  However, a similar
effect may be seen for X-ray-irradiation models if older CTTSs are
more X-ray luminous than younger CTTSs.

The [\ion{O}{1}] emission from TW Hya may be produced by recombination
subsequent to photoionization.  The [\ion{O}{1}] lines
 are centered at the
stellocentric velocity, do not correlate with optical veiling
\citep{Ale02}, and are not seen from WTTSs.  
[\ion{O}{1}] emission is thought to arise
from the disk surface of some Herbig Ae/Be stars
\citep{Boh94,Ack05}.  [\ion{O}{1}] emission produced subsequent to
OH dissociation by FUV emission from the central star will be
undetectable from disks around CTTSs \citep{Sto00}.  However,
 models of EUV or X-ray irradiated disks may explain some [\ion{O}{1}]
 emission in the low-velocity components of CTTSs \citep{Fon04}.
That the [\ion{O}{1}] lines are
narrower than the [\ion{Ne}{2}] line suggests that, if the [\ion{O}{1}]
emission is
produced in a disk, it traces less turbulent gas or gas at larger disk
radius than the [\ion{Ne}{2}] emission.

High-resolution spectroscopy of [\ion{Ne}{2}] emission from TW Hya
suggests formation 
in a disk irradiated by X-ray or EUV photons, although formation in a
heated photosphere cannot be ruled out.  The origin of the line
broadening and
the ionization path can be identified by high-resolution spectra of
additional sources.  If the line broadening is dominated by Keplerian
rotation, the [\ion{Ne}{2}] line will appear much broader from stars
with disks viewed at higher inclinations.   If the line broadening is
dominated by turbulence, the observed line width from disks viewed at
higher inclinations will only be modestly larger.
 If formed in a photoevaporative wind, the
[\ion{Ne}{2}] emission from stars with optically-thick inner disks
will likely appear blueshifted, as only the front side of the disk
will be detectable.  The description of [\ion{Ne}{2}] emission from TW
Hya should apply generally to stars
with small mass accretion rates.  [\ion{Ne}{2}] emission from stars
with large mass accretion rates may also be produced in shocks
associated with outflows.

\section{Acknowledgements}
We thank the anonymous referee  for helpful comments that improved the clarity of the discussion. This observing time was awarded by Caltech as part of the Keck/Gemini exchange program.  
We thank Kevin Volk for observing support with MICHELLE during the run and for helping to identify the atmospheric lines with ATRAN.  We thank Andrew Pickles for observing TW Hya during the re-commissioning of the East Arm Echelle spectrograph.  We thank David Hollenbach for discussing the application of photoevaporation models to the [\ion{Ne}{2}] line.


\clearpage

\begin{table}
\label{tab:obsx.tab}
{\scriptsize
\begin{tabular}{ccccccccccccc}
\multicolumn{13}{l}{Table 1: Source Properties}\\
\hline
Star   & SpT &  d & $v_r$ & $L_X$  &$M_*$ &$R_*$  & $\dot{M}$          & $F_{12.8}^a$        & $S/N^b$ & EW ([\ion{Ne}{2}])  & $L_{Ne II}$ & Refs\\
  & & pc & \kms\ & $10^{-4} L_\odot$  & $M_\odot$ & $R_\odot$ & $10^{-9} M_\odot$ yr$^{-1}$  & & &\AA & $10^{-6} L_\odot$ & \\
\hline
TW Hya & K7 & 51  & 12.2 & 1.4$^c$  & 0.7 & 1.0 &  $1.8,0.5^d$   & $0.95$  &  1.3 &$62\pm11$ & $4.8$  &1,2,3,4\\
BP Tau & K7 & 140 & 15.8 & 1.8$^e$  & 0.5 & 2.0 & $24$     & $0.7$   &  1.5 &$<10$   & $<4$ &5,6,7,8 \\
DP Tau & M0 & 140 & 17$^f$ & 0.8$^e$ & 0.5 & 1.4 & $9.0$    & $1.0$   &  1.8 &$<11$  & $<7$ &5,6,7,8\\
\hline
\multicolumn{13}{l}{$^a$Continuum flux level ($10^{-15}$ erg cm$^{-2}$ s$^{-1}$ \AA$^{-1}$) at 12.8$\mu$m from {\it Spitzer} IRS spectra}\\
\multicolumn{13}{l}{$^b$S/N per pixel in the continuum in our MICHELLE observations}\\
\multicolumn{13}{l}{$^c$0.45--2.25 eV luminosity from high-resolution {\it XMM} spectra \citep{Stel04}.}\\
\multicolumn{13}{l}{$^d$1.8 from \citet{Ale02} and \citet{Her04}, adjusted for 51 pc distance.  0.5 from \citet{Muz00}}\\
\multicolumn{13}{l}{$^e$0.3--10 eV from models based on low-resolution {\it XMM} spectra \citep{Brig07}.}\\
\multicolumn{13}{l}{$^f$Average radial velocities of GO Tau and V955 Tau from \citet{Her95}.}\\
\multicolumn{3}{l}{$1$\citet{Mam05}} &
\multicolumn{3}{l}{$2$\citet{Uch04}}  &
\multicolumn{3}{l}{$3$\citet{Wei00}} &
\multicolumn{3}{l}{$4$\citet{Web99}}\\
\multicolumn{3}{l}{$5$\citet{Val93}} &
\multicolumn{3}{l}{$6$\citet{Her95}} &
\multicolumn{3}{l}{$7$\citet{Brig07}}&
\multicolumn{3}{l}{$8$\citet{Har98}}\\
\end{tabular}}
\end{table}



\clearpage

\begin{table}
{\scriptsize \begin{tabular}{cccccccccc}
\multicolumn{10}{l}{Table 2: Selected Emission Lines from TW Hya, in order of FWHM$^{a,b}$}\\
\hline
& Line    & $\lambda_{obs}$ & $E^\prime$ & R   & $v$ & FWHM$^a$  & EW
& Flux$^c$ & Formation$^d$\\
& ID            & \AA    & eV &$\lambda/\Delta\lambda$ & \kms & \kms & \AA\ & &\\
\hline
& H$_2^e$         & 21218     & 0.60 & 49000 &  0   & $<6$ & 0.02 & 1 & D\\
& \multicolumn{2}{c}{CO (fundamental)$^f$} & & 24000 &   0  & 8.3  & -  & - & D\\
& \ion{Si}{1}   & 2882.551    & 5.1 & 30000 &   1  & 11 & 0.14 & 0.96 & P\\
& [\ion{O}{1}]$^b$ & (5577.571)  &2.0 & 36000 & (0)  & (12) & (0.11) & $<1.4$ & --\\
&  [ \ion{O}{1}] & 6300.549 &2.0 & 34000 & -1    & 12 & 0.67 & 8.7 & D?\\
& [\ion{O}{1}] & 6364.050 &2.0& 26000 &  0   & 13 & 0.23 & 3.5 & D?\\
& [\ion{S}{2}]$^b$  & (6730.81)      & 1.8    & 20500.& (0)  & (12) & $<0.06$ & $<0.9$ & --\\
& [\ion{N}{2}]$^b$  & (6583.46)      &  1.9   & 21000 & (0)  & (12) & $<0.07$ & $<1.0$ & --\\
& [\ion{Fe}{2}]$^b$  & (8616.952) &  1.7   & 27500 & (0)  & (12) & $<0.07$ & $<1.2$ & --\\
& \ion{Na}{1}$^g$          & (5889.951)  & 2.1 & 22000 & (0)  & (12) &
$<0.45$ & $<4.5$ & --\\
& \ion{Na}{1}$^g$          & (5895.924)  & 2.1 & 22000 & (0)  & (12) &
$<0.45$ & $<4.5$& --\\
& H$_2^h$         & FUV$^h$    & 1--2$^i$ & 25000 &   0  & 14 & -  & - & D\\
&  \ion{Fe}{2}  & 2371.319  & 5.6 & 30000 &   -1  & 15 & 0.17 & 0.76 & ?\\
&  \ion{Fe}{2}  & 2665.569  & 8.0 & 30000 &   0  & 15 & 0.09 & 0.55 & ?\\
& \ion{Ca}{2}   & 8542.362     & 3.2 & 16000 &  -2   & 15 & 1.2 & 21 & P+A?\\
& \ion{Ca}{2}   & 8662.464   & 3.1& 43000 &  -1    & 16 & 0.84   & 17 & P+A?\\
& \ion{Mg}{2}   & 2791.722   & 8.9 & 30000 &  1   & 16 & 0.27 & 0.19 & P?\\
& {\bf [\ion{Ne}{2}]} & {\bf 128140}       & {\bf 0.10} & {\bf 31000} & {\bf -2} & {\bf 22.5}  & {\bf 62}   & {\bf 5.9} & D\\
&  \ion{Fe}{2} (1)$^j$  & 2626.565  & 4.8 & 30000 &   1  & 25 & 1.3 & 6.3 & P/H?\\
& (2)$^j$           & 2626.557    & 4.8 & 30000 &   0  & 130 &2.1 & 10.3 & H?\\
& \ion{Cl}{1}   & 1351.715    & 9.3  & 30000 &  0 & 41 & 0.18  & 0.27 & H?\\
& \ion{C}{1}   & 1656.346    & 7.5  & 30000 &  1   & 41 & 1.5  &4.6 & H?\\
&\ion{C}{2}]   & 2326.231    & 5.3 & 30000 &   1 & 42  &3.8 & 19 & H?\\
& \ion{He}{1}$^k$   & 5876.060    & 23.1 & 19000 &  10  & 50 & 0.81 & 12 & A\\
& \ion{S}{1}   & 1296.231    & 9.6  & 30000 &  0     & 52 & 0.95  &1.3 & A?\\
& \ion{N}{1}]   & 1411.966    & 12.4  & 30000 &  -7    & 83 & 1.9 & 2.8 & A?\\
& \ion{He}{2}$^l$ & 1640.601   & 48  & 30000 & 20  & 88 & 310  & 94 & A\\
& \ion{O}{3}]   & 1666.093    & 7.5  & 30000 &  -23  & 113 & 0.8  &1.6 & W?\\
& \ion{C}{3}     & 2297.842    & 18.1 & 30000 &  22  & 127 & 0.8 & 3.2 & A\\
& H$\gamma$     & 4341.214    & 13.1 & 32000 &  39  & 195 & 8.1 & 81 & A\\
& H$\beta^m$     & 4861.713   & 12.8 & 32000 &  11  & 199 & 34 & 44 & A\\
& H$\delta$   & 4102.464    & 13.2 & 32000 &  41  & 201 & 2.9 & 23 & A\\
& \ion{C}{4}$^l$ & 1548.757   &  8.0  & 30000 &  110 & 300 & 93 & 186 & A\\
& \ion{N}{5}$^l$ & 1239.068   & 10.0  & 30000 &  60  & 340 & 15 & 30 & A\\
\hline
\multicolumn{10}{l}{$^a$Intrinsic FWHM after deconvolution from the instrumental resolution.}\\
\multicolumn{10}{l}{$^b$Assumed properties for undetected lines in parentheses.}\\
\multicolumn{10}{l}{$^c$$10^{-14}$ erg cm$^{-2}$ s$^{-1}$}\\
\multicolumn{10}{l}{$^d$W=Wind, P=photosphere, H=Heated Photosphere, A=Accretion funnel, D=Disk}\\
\multicolumn{10}{l}{$^e$From Weintraub et al. (2000).}\\
\multicolumn{10}{l}{$^f$Averaged CO FWHM from Salyk et al. (2007).}\\
\multicolumn{10}{l}{$^g$Not detected here.  \citet{Ale02} found narrow
\ion{Na}{1} D lines at the stellocentric velocity.}\\
\multicolumn{10}{l}{$^h$Coadded FUV H$_2$ profiles by coadding many of the strongest lines from Herczeg et al. (2002).}\\
\multicolumn{10}{l}{$^i$Excited from many levels between 1--2 eV (Herczeg et al. 2002).}\\
\multicolumn{10}{l}{$^j$NUV \ion{Fe}{2} lines from TW Hya have narrow
  and broad components.}\\
\multicolumn{10}{l}{$^j$Properties are for bright, narrower
  component.  Excess weak redshifted emission is also present.}\\
\multicolumn{10}{l}{$^l$Line center and width calculate from a Gaussian fit to a non-Gaussian profile, see Johns-Krull \& Herczeg (2007).}\\
\multicolumn{10}{l}{$^m$Corrected for optically-thin wind absorption}\\
\end{tabular}}
\end{table}

\clearpage

\begin{figure}
\plotone{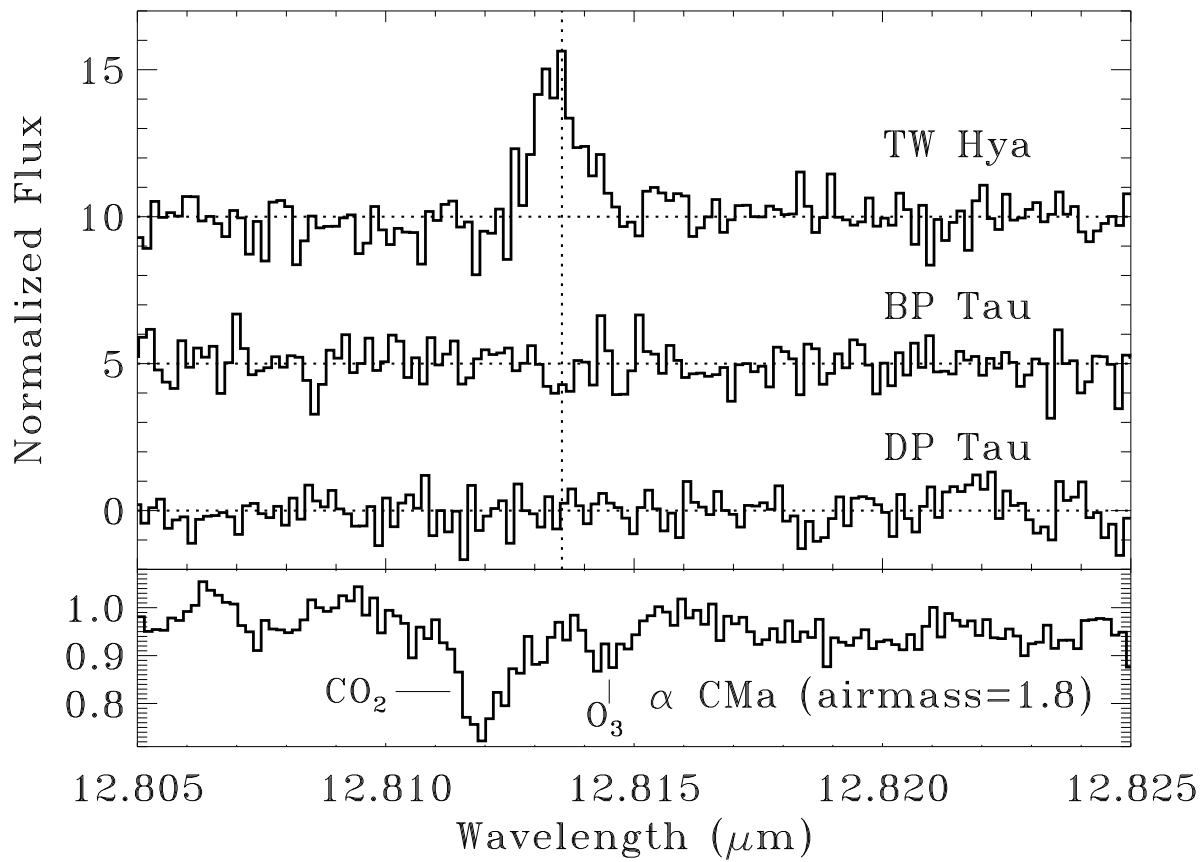}
\caption{The MICHELLE spectra of TW Hya, BP Tau, and DP Tau, shifted to the appropriate stellar velocity.  The [\ion{Ne}{2}] line is strong from TW Hya but not detected from BP Tau or DP Tau.  The mid-IR continuum is detected from all three sources.  The bottom panel shows the spectrum of the telluric standard $\alpha$ CMa shifted to the observed velocity of TW Hya.}
\end{figure}

\clearpage

\begin{figure}
\plotone{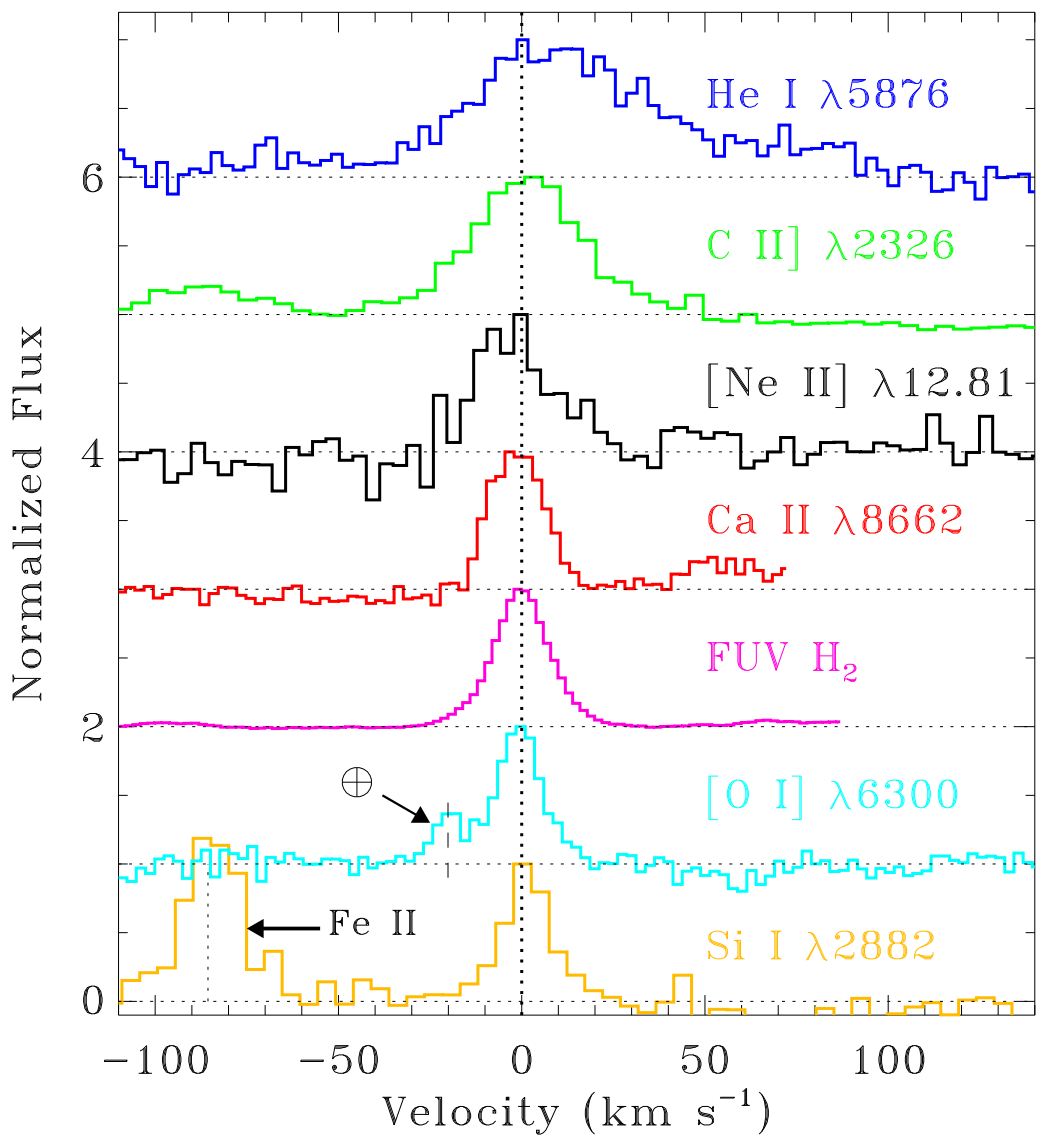}
\caption{A comparison of the [\ion{Ne}{2}] line profile with many
  optical and UV lines from TW Hya.  The [\ion{Ne}{2}] line is much
  more narrow than lines associated with the accretion funnel
  (including the \ion{He}{1} $\lambda5876$ line shown here), but is broader than
  lines formed in the disk (H$_2$ and possibly [\ion{O}{1}] emission
  shown here) and in the photosphere (\ion{Si}{1} emission shown here).  Telluric [\ion{O}{1}] emission is identified by the Earth symbol.}
\end{figure}

\end{document}